# Generating extreme electric fields in 2D materials by dual ionic gating

Benjamin I. Weintrub[1], Yu-Ling Hsieh[1,2], Jan N. Kirchhof[1], and Kirill I. Bolotin*[1]

1. Department of Physics, Freie Universität Berlin, Berlin, Germany
2. Department of Mechanical Engineering, National Central University, Taoyuan City, Taiwan

**Abstract**

We demonstrate a new type of dual gate transistor to induce record electric fields through two-dimensional materials (2DMs). At the heart of this device is a 2DM suspended between two volumes of ionic liquid (IL) with independently controlled potentials. The potential difference between the ILs falls across an ultrathin layer consisting of the 2DM and the electrical double layers above and below it, thereby producing an intense electric field across the 2DM. We determine the field strength via i) electrical transport measurements and ii) direct measurements of electrochemical potentials of the ILs using semiconducting 2DM, WSe$_2$. The field strength across the material reaches more than 3.5 V/nm, the largest static electric field through any electronic device to date. We demonstrate that this field is strong enough to close the bandgap of trilayer WSe$_2$ driving a semiconductor-to-metal transition. Our approach grants access to previously-inaccessible phenomena occurring in ultrastrong electric fields.

**Introduction**

Electric fields are widely used to control material properties and to explore new and diverse physical phenomena. The first group of phenomena appears due to changes of the carrier density induced in a material by the field at its surface, typically explored using field-effect transistors (FETs)[1,2]. Electric fields also cause a second, qualitatively different, group of effects when the field penetrates through the material's bulk. In this case, the presence of a field inside the material breaks symmetries[3–7], bends the band structure along the direction of the field[3,5–7], and modifies the energetics of excitons with a dipole moment parallel to the field[3–6,8,9]. In conventional FETs, the induced carriers at the material's surface almost completely screen the field in the bulk of the material. Therefore, a dual gate FET with a pair of gate electrodes above and below the material under study is used to explore the effects of an external electric field penetrating a material. In this configuration, the field strength is controlled by the potential difference between the bottom and top gates, while the Fermi level is determined by their sum[5,7–9]. However, an important limitation for studying large electric fields in conventional solid-state FETs is the breakdown of gate dielectrics happening at around 0.5 – 1.0 V/nm[10–16] (somewhat larger dielectric strengths are measured using local probe techniques[11,12]).

The limitation on the maximum achievable carrier density has been overcome during the last decade via ionic gating, which combines condensed matter physics with electrochemistry[17,18]. In that technique, an ionic compound such as ionic liquid (IL), a molten salt, is placed over a material under study[17–20]. A potential applied between the gate electrode inside the liquid and the sample falls predominantly over an atomically thick (≤ 1 nm) electric double layer (EDL) at the IL/sample interface, modeled as a capacitor with an exceptionally large geometric areal capacitance (~10 µF/cm$^2$)[17,19,21–27]. The resulting electric field inside the EDL induces a carrier density inside the material[27,28]. Critically, the field generated here is not limited by the dielectric breakdown of gate dielectrics, which limit the performance of conventional solid-state FETs. Instead, the only significant limitation is electrochemical modification of the material or electrodes, which occurs when the potential drop across a particular interface is too large (outside the electrochemical window)[27,29]. Ionic gating enabled previously inaccessible carrier densities larger than 10$^{14}$ cm$^{-2}$ to be reached[19,28,30,31]. The interactions between electrons at these carrier densities result in structural phase transitions[32] and new electronic phases such as exotic superconductivity[30,31] and gate-controlled ferromagnetism[33,34]. These effects are especially pronounced in 2DMs such as graphene, transition metal dichalcogenides (TMDCs), or phosphorene, where the carriers are spatially confined to one or few atomic layers[18].

Despite the progress in using ionic gating to induce high carrier densities, dual ionic gating has not been used to generate intense external electric fields inside the bulk of materials. Although single-gated suspended 2DMs[35,36] as well as hybrid dielectric/ion approaches to dual gating[37–39] have been used, no ionic counterpart to dual gate FETs has been demonstrated. Because of that, a wide range of phenomena predicted to emerge at fields stronger than $F_\perp \sim 1$ V/nm remains inaccessible in solid state devices. For example, for fields near $F_\perp \sim 2 - 3$ V/nm, the interlayer bandgap of bilayer (2L) TMDCs is expected to decrease to zero[40,41]. In this situation, interlayer excitons should start forming at zero energy costs and a transition into a new state of matter, an interlayer excitonic insulator[42], may occur. Other predicted yet unobserved phenomena at extreme fields include an insulator to topological insulator transition in phosphorene ($F_\perp > 3$ V/nm)[43], topological insulator to semimetal to normal insulator transition in 1T' TMDCs ($F_\perp > 2$ V/nm)[44], structural change in chirality for monolayer Te ($F_\perp > 7$ V/nm)[45], and giant valley polarization ~65 meV in WSe$_2$/CrSnSe$_3$ heterostructures ($F_\perp \sim 6$ V/nm)[46].

Here, we develop a double-sided ionic gating approach to generate ultrastrong electric fields. The approach can be viewed as a counterpart to conventional dual gated FETs which is not limited by the breakdown of gate dielectrics. To generate the field inside a 2DM, we apply a potential difference between the two ILs above and below the 2DM, thereby generating a new type of EDL consisting of two ILs separated by an ultrathin membrane. We use a combination of electrochemical and electrical transport measurements to record an electric field of at least 3.5 V/nm inside 2DMs, more than three times larger than the biggest fields reported for conventional solid-state FET technologies.

**Device concept**

At the core of our approach to generate and measure large perpendicular electric fields is an electrically-contacted few-layer 2DM suspended between two volumes of IL (Fig. 1a). The potential difference between the top and bottom ILs, $\Delta V^{\text{ref}}$, is controlled by separate top and bottom gate electrodes in contact with their respective ILs. This potential difference falls across an ultrathin capacitor, thickness $d_\perp$, consisting of the 2DM (thickness $Nd_{\text{int}}$ for $N$-layer TMDCs with interlayer spacing[47] $d_{int} \approx 0.6$ nm) and two EDLs (average thickness[21–26] $d_{EDL} \approx 0.5$ nm), one above and one below the 2DM. Neglecting the dielectric constant variation at the nanometer scale, we estimate the field inside that layer as $F_\perp = \Delta V^{ref}/d_\perp$, where $d_\perp = Nd_{int} + 2d_{EDL}$ is the layer thickness. Approximating $\Delta V^{ref} \approx 6$ V (corresponding to the potential of top and bottom ILs at ±3 V, the maximum electrochemical window of our IL), we can estimate a maximum achievable field strength of ~3.8 V/nm for our 1L devices, at least three times larger than what is possible for dielectric-based devices. The maximum achievable field reduces with growing $d_\perp$.

We determine the field strength using a specific 2DM, few-layer WSe$_2$, as a field sensor. We use electrical transport measurements to determine the bandgap of the 2DM[19,27], for which multilayers are known to exhibit a strong dependence on perpendicular electric fields[9,40,41]. When no field is present, the band structure of few-layer WSe$_2$ can be approximated as energy-degenerate conduction and valence bands, one set of bands per layer. An external perpendicular electric field breaks the inversion symmetry of the 2DM, thereby inducing a maximum energy difference of $e(N - 1)d_{int}F_\perp$ between the outer layers, where $e$ is the elementary charge and $(N - 1)d_{int}$ is the distance between the centers of outer layers of the $N$-layer 2DM. Correspondingly, the bandgap of multilayer WSe$_2$ becomes spatially indirect, occurring between one layer's conduction band and another layer's valence band (Fig. 1b). We can estimate that the bandgap reduces from the field-free value, $E_N^0$, down to $E_N = E_N^0 - e(N - 1)d_{int}F_\perp$. Therefore, by extracting $E_N$ from electrical transport measurements, we can directly determine the field strength (SI section S2). The effects of free carrier screening are neglected in this analysis, because the Fermi level is inside the bandgap. Although this simplified model does not take into account e.g. hybridization between bands or screening, photoluminescence data[47] and detailed DFT calculations[48] for bilayer WSe$_2$ predict a linear dependence of interlayer bandgap on the interlayer field with $e^{-1}dE_{2L}/dF_\perp = d_{int} \approx 0.6 \pm 0.1$ nm, very close to the interlayer separation in 2L WSe$_2$.

**Device fabrication**

A schematic of the dual IL gated FET is shown in Fig. 2a. We choose WSe$_2$ due to its relatively small bandgap[49–51]. We suspend WSe$_2$ over a ~10 μm$^2$ rectangular hole in a silicon nitride (SiN) membrane on a Si substrate. Gold electrodes on top of the SiN electrically contact the 2DM. A separate pair of electrodes on top of the SiN are used as the top gate electrode, which sets the potential of the top IL, and top reference electrode, which measures the potential of the top IL (Fig. 2b). Just before the measurement, a drop of IL (DEME-TFSI[20]) is placed on the back of the Si/SiN chip, which is subsequently placed onto a sapphire chip with gold electrodes serving as bottom gate/reference electrodes. Finally, a drop of IL is deposited onto the device such that it is not in contact with the lower liquid, and the entire assembly is loaded into an electrical probe station where it is measured at room temperature and low pressure, ~10$^{-5}$ mbar (Fig. 2c, SI section S1).

**Bandgap determination from transport measurements in dual IL-gated devices**

We turn to transport measurements of dual IL-gated FETs. We record a map of $I_{ds}$ vs. ($V_b$, $V_t$) for a bilayer (2L) WSe$_2$ device, sample #1 (Fig. 3a). Assuming equal coupling of top and bottom ILs to the material, the Fermi level of the system is controlled by $V_g = V_b + V_t$, whereas the field across the 2DM depends on $\Delta V = V_b - V_t$ (SI section S2). In constant-field traces, $I_{ds}(V_g, \Delta V = \text{const})$, we observe ambipolar transport with a region of negligibly low current (Fig. 3b). This region corresponds to the Fermi level ($E_F$) inside the bandgap of the material, while the areas of conductance to the left or right of that region correspond to $E_F$ within the valence or conduction band respectively[5,9,19,33,52]. When the field is increased ($\Delta V > 0$ V) the region of zero current in Fig. 3b shrinks. That behavior is particularly clear in Fig. 3a where the region of zero conductivity (black region) along the $V_g$ direction gradually reduces with increasing $\Delta V$. Quantitatively, the bandgap is calculated from transport measurements as $E_{2L} = \frac{1}{2}e\alpha(V_e - V_h)$, where $V_e$ and $V_h$ are conduction and valence band threshold voltages respectively (which correspond to the Fermi level at the conduction and valence band extrema)[27], and $\alpha \approx 0.77$ is the gating efficiency (SI section S2). The bandgap determined in this manner decreases with $\Delta V$ and is linearly dependent on it (Fig. 3c). From the linear fit to this graph, we find that the bandgap reduces to zero at $\Delta V_{close} \approx 5.9$ V. Using $\Delta V_{close} = 2E_{2L}^0(d_{int} + d_{EDL})/(e\alpha d_{int})$, we can experimentally determine the thickness of the EDL from this data. We obtain $d_{EDL} \approx 0.2$ nm, smaller than the value of ~0.5 nm quoted in the literature[21–26]. This discrepancy could stem from the redistribution of charges inside the ions or signal the limitations of our simplified model.

Next, we note that in our geometry the same $\Delta V$ should produce a larger bandgap reduction in a few-layer 2DM compared to a bilayer. This is simply the result of a larger fraction of $\Delta V^{ref}$ falling across the 2DM compared to the EDLs as $N$ increases (SI section S2). We therefore measured a 3L WSe$_2$ device (sample #2, Fig. 4). In contrast to the 2L device, we observe bandgap closing at around $\Delta V \approx 4$ V, well within the electrochemical window. Therefore, a stronger response of that system to $\Delta V^{ref}$ allows the observation of a field-driven semiconductor-to-metal transition.

**Quantifying field strength**

Finally, we turn to the central result of this paper, determination of field strength in our devices. For the trilayer device, the field strength corresponding to bandgap closing is $F_\perp = E_{3L}^0/(2ed_{int}) \approx 1.4$ V/nm (where we use $E_{3L}^0 = 1.7$ eV). For the bilayer data of Fig. 3, we can determine the field as $F_\perp = (E_{2L}^0 - E_{2L})/(ed_{int})$ where we recall that $E_{2L}$ depends on $V_e - V_h$ (Fig. 5b, black points). The linear fit to these datapoints indicates that the largest field achieved in that device at our maximum $\Delta V = 5$ V is $F_\perp \approx 2.5 \pm 0.3$ V/nm (Fig. 5b, black dashed line). We expect that the smallest $d_\perp$, and hence the largest $F_\perp$, should be reached for monolayers. Unfortunately, the bandgap of monolayer WSe$_2$ does not depend on $F_\perp$. We experimentally confirmed this expectation in our approach (Fig. S3 in SI). Therefore, transport measurements cannot be used to determine the field. Instead, we rely on electrochemical measurements. For 1L WSe$_2$ (sample #3), we sweep a potential difference between bottom and top gate electrodes, $\Delta V = V_b - V_t$, while recording the potentials of top and bottom ILs, $V_t^{ref}$ and $V_b^{ref}$ respectively, as measured by corresponding reference electrodes. The potential dropping across the 2DM and the two EDLs, $\Delta V^{ref} = V_b^{ref} - V_t^{ref}$, depends on $\Delta V$ linearly (Fig. 5a). The hysteresis seen in such data is commonly seen in ionic gating and stems from the delayed response of ions. We calculate the electric field strength across the 2DM as $F_\perp = \Delta V^{ref}/d_\perp$ (Fig. 5b, red line). The field determined in this fashion reaches ~3.5 ± 0.2 V/nm near the limit of our electrochemical window using the literature value for the EDL thickness $d_{EDL}$ = 0.5 ± 0.1 nm. We consider this a conservative estimate, as the field is even larger using $d_{EDL}$ estimated from transport measurements. The fields we reach in these devices are the largest static electric fields ever reported to date through any electronic device, to the best of our knowledge. This measured field surpasses the dielectric strength of common gate dielectrics such as hBN[10,11], $SiO_2$[12], SiN[13,14], and $HfO_2$[15].

**Discussion**

The largest field strength in our measurements, greater than 3.5 V/nm, should already be sufficient to induce multiple predicted but still unobserved electronic and structural phases in various 2DMs[40,41,43,44]. The maximum field strength we report in 2L $WSe_2$ is limited by the electrochemical window of the IL we use. Conversely, the maximum field we report in 3L $WSe_2$ is limited by the material we use to record the field strength. The bandgap of 3L $WSe_2$ closes at $F_\perp \approx 1.4$ V/nm, preventing measurements of higher field strengths using transport measurements. Even larger fields can generally be achieved by reducing $d_\perp$ (using thinner 2DMs and/or using ILs with smaller double layer thickness), increasing $\Delta V^{ref}$ (using ILs with larger electrochemical windows), or using large-bandgap materials to measure the field. It is interesting to note that the fields extracted from our measurements are on the same scale or higher compared to the dielectric strength of 2DMs and common dielectrics. While the maximum field in dielectric-based devices is limited by the breakdown of both the gate dielectric as well as the material under study, in ionic gating approaches the limitation is due to electrochemistry at interfaces. Electric currents across the IL/2DM interface can only flow when the potential difference across that interface exceeds its electrochemical window. We finally mention that our device provides the first demonstration of a new type of EDL consisting of IL/2DM/IL instead of the conventional IL/metal electrochemical system. This novel EDL may unlock new areas of research e.g. dielectric breakdown of materials in extreme fields which could lead to enhanced FET gate dielectrics.

**Summary**

In conclusion, we developed a new type of FET allowing dual electrochemical gating of 2DMs or other ultrathin films. We demonstrated that our technique can reach electric fields greater than 3.5 V/nm inside 2DMs through a pair of independent measurements: electrochemical reference voltage measurements and electrical transport measurements. We also observe at least ~1.4 V/nm through trilayer $WSe_2$, which drove a field-induced semiconductor-to-metal transition where electrons and holes are localized on opposite outer layers of the trilayer. To the best of our knowledge, we reach the highest static external electric field penetrating through the entire bulk of a material in a condensed matter system. Our dual IL-gated design should be compatible with a large variety of 2DMs, other thin films, ILs, and other ionic compounds (such as ionic gels, polymer electrolytes, etc…)[17]. We expect that the ability to induce very large fields in condensed matter systems will unlock access to novel predicted but unobserved states of matter, and it is also worth noting that our experimental geometry may allow increasing, by a factor of two, the total gate-sample capacitance achievable in conventional ion-gated FETs with just one gate due to both sides of the 2DM being exposed to ILs in our devices[35,36]. Finally, our novel dual IL gate FET design should facilitate exploration of new areas of research requiring ultrahigh carrier densities and/or ultrahigh static electric fields previously experimentally inaccessible in condensed matter systems.

**Acknowledgments**

We gratefully acknowledge Dr. Kyrylo Greben for useful discussions. This work was supported by the Deutsche Forschungsgemeinschaft (DFG)—Projektnummer 182087777—SFB 951 and ERC Starting Grant No. 639739.

**Author contributions**

BIW, YH, and JNK fabricated the samples. BIW and YH conducted the measurements. BIW analyzed the data and carried out the simulations. KIB conceived the approach and supervised the project. All authors contributed to writing the manuscript.

**Competing interests**

The authors declare no competing interests.

**Data availability**

The data that support this study are available from the authors upon reasonable request.

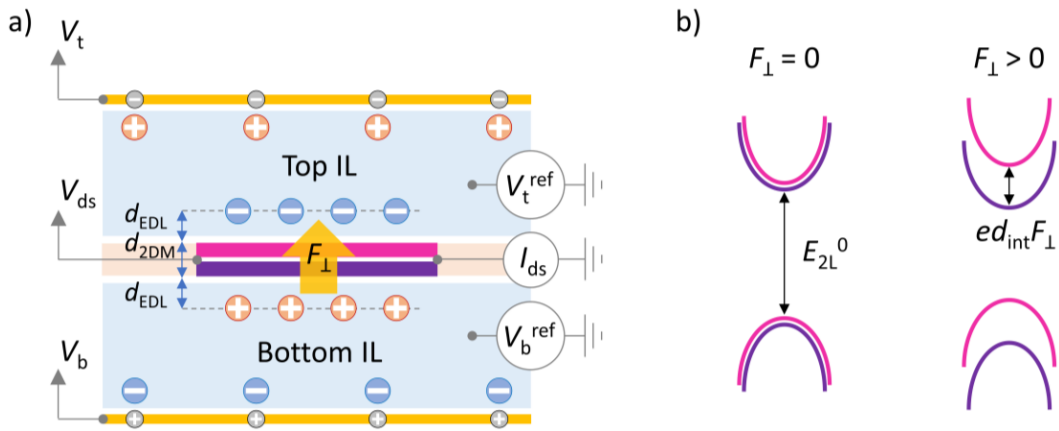

**Fig. 1 | Dual ionic liquid gating. a)** Concept of a dual ionic liquid-gated bilayer 2DM device. The potentials of the top and the bottom ionic liquids are independently controlled by top and bottom gate voltages, $V_t$ and $V_b$ respectively. The potential difference between top and bottom ionic liquids drops over a ~2 nm thick layer (including the bilayer), thereby generating an ultrastrong electric field through the bilayer. **b)** A band structure sketch of a bilayer 2DM at zero and non-zero electric field. The field breaks the degeneracy between the energy bands corresponding to opposite layers and reduces the overall bandgap by $ed_{int}F_\perp$.

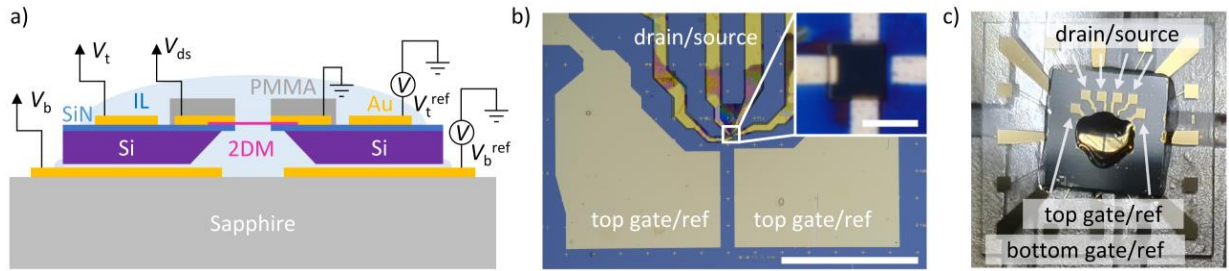

**Fig. 2 | Device and measurement overview. a)** Side-view cartoon of the device and measurement scheme. **b)** Microscope (5x) image of sample #1 before applying ionic liquids. Drain and source electrodes are covered with crosslinked PMMA. Scale bar is 1 mm. **Inset:** 100x image of the 2DM deposited onto a square hole (~4 µm x 4 µm) in SiN. Everything in the inset other than the area of the 2DM over the hole is covered by crosslinked PMMA. Scale bar is ~4 µm. **c)** Photograph of sample #1 just before measurement (SiN chip is 5 mm x 5 mm). Notice: the top and bottom ionic liquids are not in contact.

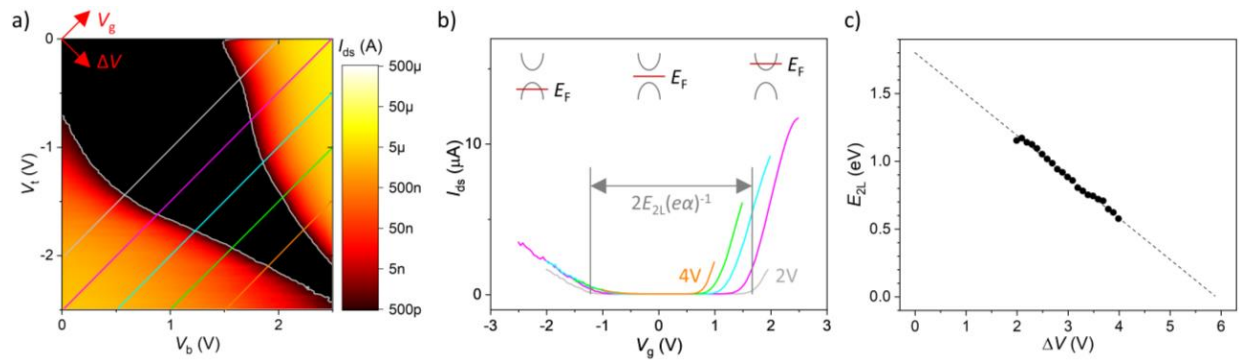

**Fig. 3 | Transport measurement of dual ionically-gated bilayer WSe$_2$. a)** Map of $I_{ds}$ vs. ($V_b$, $V_t$) for bilayer WSe$_2$ (sample #1). **b)** Line scans corresponding to the slices of $\Delta V$ from the map in *a*). The bandgap, $E_{2L}$, at each $\Delta V$ is related to the difference between threshold voltages, indicated for the gray curve. Note how the bandgap appears to close as $\Delta V$ increases. Cartoons indicate the Fermi level position relative to the band edges. **c)** Measured values of $E_{2L}$ vs. $\Delta V$ extracted from the line scans as well as a linear fit (dashed line). The bandgap of the bilayer should close around $\Delta V_{close} \approx 5.9$ V.

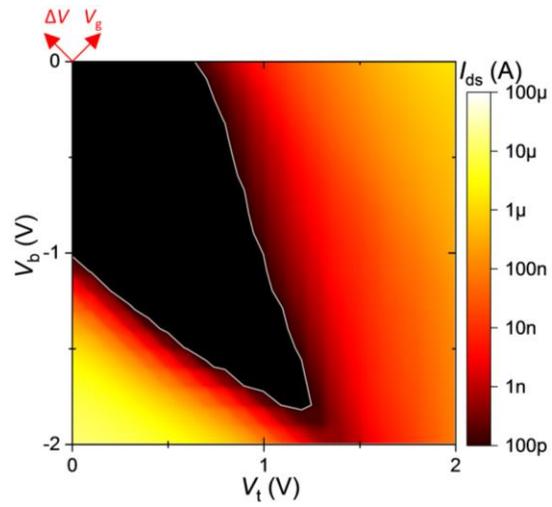

**Fig. 4 | Transport in trilayer WSe$_2$.** Map of $I_{ds}$ vs. ($V_b$, $V_t$) for 3L WSe$_2$ device (sample #2). As $|\Delta V|$ increases from 0 V, the bandgap shrinks, and it eventually closes at a critical value of $F_\perp$ around 1.4 V/nm.

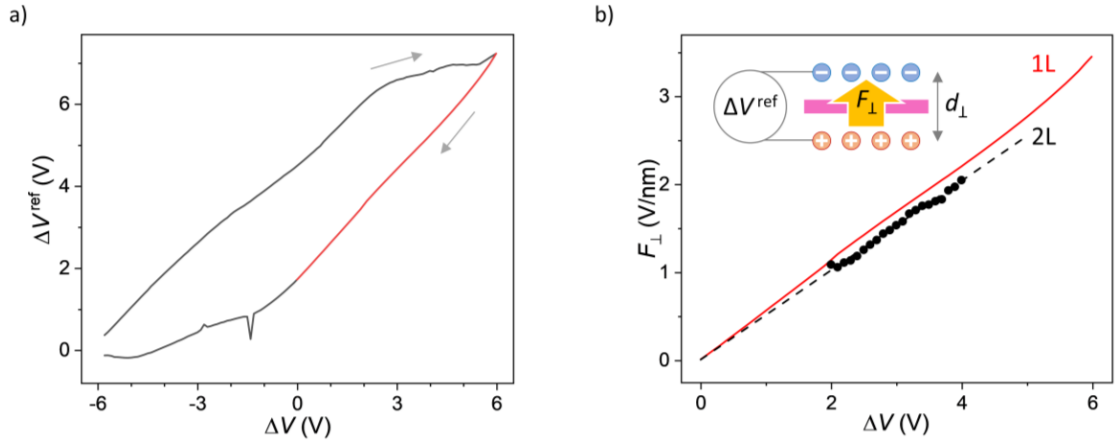

**Fig. 5 | Electric field through 2L and 1L WSe₂. a)** Reference voltage data from 1L WSe$_2$ (sample #3). The direction of the sweep is indicated by arrows, and the red portion of the curve is used to extract the field. **b)** Perpendicular electric field through 1L WSe$_2$ extracted from reference voltage data in *a*) (red line) alongside the 2L WSe$_2$ field data and fit line from Fig. 3 (black dots and dashed line).

# Supplementary information for

# "Generating extreme electric fields in 2D materials by dual ionic gating"

Benjamin I. Weintrub[1], Yu-Ling Hsieh[1,2], Jan N. Kirchhof[1], and Kirill I. Bolotin*[1]

1. Department of Physics, Freie Universität Berlin, Berlin, Germany
2. Department of Mechanical Engineering, National Central University, Taoyuan City, Taiwan

## S1. Sample fabrication details

Fig. S1 shows images of all samples used in the main text at various stages of the fabrication process. Silicon nitride (SiN) membranes were purchased from Norcada (20 nm SiN on Si). A ~3 μm x 3 μm hole in the SiN membrane was created using focused ion beam lithography. Flakes of 2D materials (2DMs) were mechanically exfoliated onto a thin piece of PDMS purchased from GelPak (~6 mils thick, X4 retention factor) using the standard "scotch tape" technique[1]. The flakes were subsequently dry transferred[2] over the hole in the SiN membrane at 60 °C using a homemade transfer system. We fabricate electrodes using standard electron beam lithography (EBL), thermal evaporation (3 nm Cr and 70 nm Au), and liftoff (60 °C acetone for 1 – 3 hours). Following liftoff, we confirm that the suspended region of the 2DM is not ruptured. Fabrication for sample #2 in the main text ends here.

As a final step for samples #1, #3, and #4, we cover non-suspended regions of the 2DM and the drain/source electrodes with a layer of cross-linked poly(methyl methacrylate) (PMMA) EBL resist. This capping layer serves several functions (Fig. S6): 1) it restricts the FET channel area for the top IL to match the channel area for the bottom IL (just the suspended region), thereby eliminating the conductivity contribution from supported regions of the 2DM, 2) it clamps the 2DM at its edges, thereby preventing intercalation of ions in the IL to go between the layers of the 2DM, and 3) it reduces the influence of the drain/source metal electrodes on the top IL's gating efficiency. To fabricate this capping layer, we spin a layer of PMMA and a layer of conductive polymer (Electra 92 from Allresist) after the electrode fabrication step. We then perform EBL for the PMMA crosslinking step to cover the 2DM's supported region and drain/source electrodes. For our crosslink step, we use a dose of 10,500 μC/cm$^2$ at 30 kV near the suspended 2DM, and to cover the drain/source electrodes away from the 2DM we use a dose of 17,500 μC/cm$^2$ at 10 kV. Lastly, we remove the excess unexposed PMMA in heated (60 °C) acetone for 3 hours. Fabrication for samples #1 and #4 in the main text ends here. To fabricate sample #3 from the main text, we performed this crosslinking step in a 1 mm x 1 mm area around the 2DM immediately following the dry transfer step, and then removed the excess PMMA using the same liftoff recipe as described above.

## S2. Extracting gating efficiency, bandgap, and electric field

Our 2DM is exposed to two individual ILs, one above and one below the 2DM. Only a fraction of the potential applied to the gate electrodes ($V_b$, $V_t$) inside these liquids falls across the 2DM/IL interface ($V_b^{ref}$, $V_t^{ref}$), which is characterized by gating efficiencies: $\alpha_b \equiv V_b^{ref}/V_b$ and $\alpha_t \equiv V_t^{ref}/V_t$.

These gating efficiencies depend mainly on the size of the gate electrodes (relative to the 2DM and exposed drain/source electrodes). In our transport devices, we try to equalize these parameters between top and bottom gate electrodes. The size, material composition, and placement of the gate electrodes above and below the 2DM are roughly equal. Because of this symmetry between top and bottom gate electrodes, we assume $\alpha_b = \alpha_t = \alpha$. Similarly, we can relate ($V_g$, $\Delta V$) to the fraction of those applied voltages, ($V_g^{ref}$, $\Delta V^{ref}$), and when we assume $\alpha_b = \alpha_t = \alpha$, we see that $\alpha = V_g^{ref}/V_g = \Delta V^{ref}/\Delta V$. Therefore, everywhere in the main text, we just use a single parameter, $\alpha$, without specifying which gate it refers to.

We can extract $\alpha$ directly from reference voltage measurements ($\Delta V^{ref}$ vs. $\Delta V$ data in Fig. 5). We form a linear fit and note that the slope is $\Delta V^{ref}/\Delta V = \alpha \approx 90\%$ for sample #3 in Fig. 5 (red curve). We also see in Fig. S2 that the voltage drops across both gate EDLs as a function of $\Delta V$ are almost the same, confirming that $\alpha_b \approx \alpha_t$ in this device. In samples of Fig. 3 and 4, we do not record references voltages simultaneously, so instead we extract $\alpha$ from the transport data (see below).

The position of the Fermi level and gate voltages in a dual gate system are related by ½$e(\delta V_b^{ref} + \delta V_t^{ref}) = \delta E_F + e\delta\varphi$, where $\delta E_F$ is a change in Fermi energy, $\delta\varphi$ is a change of the electrostatic potential, and the factor of ½ describes the 2DM seeing the averaged chemical potentials of two equivalent top/bottom ILs. Due to the exceptionally large areal capacitance of ILs causing midgap states to fill with little applied gate voltage, when the Fermi level is in the bandgap of the 2DM (sample is charge neutral), we can neglect the $\delta\varphi$ term, and substituting $V_b^{ref} + V_t^{ref} = V_g^{ref} = \alpha V_g$, we get $\delta E_F = ½e\alpha\delta V_g$. If we now consider the change in Fermi energy to equal the bandgap energy, $E_N$, we arrive at $E_N = ½e\alpha(V_e - V_h)$, where $V_e$ and $V_h$ are electron and hole threshold voltages respectively (measured in the $V_g \equiv V_b + V_t$ domain).

Experimentally, we extract $V_e$ and $V_h$ from transport measurements by fitting a line to the electron and hole linear conduction regions of the $I_{ds}$ vs. $V_g$ data, and then we determine the value of $V_g$ where the fit goes to zero. We extract threshold voltages like this at several values of $\Delta V$ and subtract the threshold voltages. We form a linear fit to the $V_e - V_h$ vs. $\Delta V$ data, and this linear fit can be used to determine both $\alpha$ and the point at which the bandgap closes, $\Delta V_{close}$. To calculate $\alpha$, we use the value of $V_e - V_h$ when $\Delta V = 0$ V, and note that the bandgap in this case is equal to the zero-field bandgap, $E_N^0$, which leads to $\alpha = 2E_N^0 e^{-1}(V_e - V_h)^{-1}$, which we use to calculate bandgap.

To extract the field from the measured bandgap of bilayer 2DMs we use $E_{2L} = E_{2L}^0 - ed_{int}F_\perp$, where $E_{2L}^0$ is the unperturbed bandgap of the bilayer and $d_{int}$ is the distance between centers of adjacent layers in the bilayer. This equation is readily generalized to $N$-layer 2DMs by $E_N = E_N^0 - ed_cF_\perp$, where $d_cF_\perp$ is the potential difference between centers of the outermost layers of the $N$-layer 2DM, and $d_c = (N - 1)d_{int}$ is the distance between centers of the outermost layers. We can relate the field-dependent bandgap to transport measurements via $E_N = ½e\alpha(V_e - V_h) = E_N^0 - ed_cF_\perp$, and through this we can calculate $F_\perp$ as a function of threshold voltage difference as $F_\perp =$ $(ed_c)^{-1}(E_N^0 - E_N)$, which we use in the main text to analyze transport data from 2L WSe$_2$ (sample #1).

The formalism above suggests that the bandgap of monolayer WSe$_2$ does not depend on perpendicular field. In a control measurement, we fabricated a test device, sample #4, and tested this assumption (Fig. S3). As expected, we find no systematic dependence of bandgap on $\Delta V$ that is not caused by electrolyte instability.

Finally, we simulated the field-dependent transport through 1L, 2L, and 3L double-gated devices. We estimated threshold voltage difference as a function of $\Delta V$ by plugging $F_\perp = \Delta V^{\text{ref}}/d_\perp = \alpha \Delta V/d_\perp$ into $\frac{1}{2}e\alpha(V_e - V_h) = E_N^0 - ed_cF_\perp$, which gives $V_e - V_h = 2(e\alpha)^{-1}E_N^0 - 2\Delta V d_c/d_\perp$. The simulated threshold voltage maps for the devices shown in Fig. S4 look very close to experimentally measured data, shown side by side for comparison. In the simulations, we use $d_{\text{int}} = 0.6$ nm, $d_{\text{EDL}} = 0.5$ nm, $E_{1L}^0 = 1.9$ eV, $E_{2L}^0 = 1.8$ eV, $E_{3L}^0 = 1.7$ eV, and ~90 % gating efficiency.

We note that the energy of the bandgap shift, $ed_cF_\perp = e(d_c/d_\perp)\Delta V^{\text{ref}}$, depends on how much of $\Delta V^{\text{ref}}$ falls across $d_c$ compared to the entire $d_\perp$ over which it is applied. Assuming that $d_{\text{EDL}} \approx d_{\text{int}}$, we find that $d_c/d_\perp \approx (N-1)/(N+2)$. This largely explains why bandgap closing can be observed within our electrochemical window for 3L, but not for 2L WSe$_2$. Since the zero-field bandgap of WSe$_2$ tends to decrease with $N$, this also aids in observing bandgap closing in 3L WSe$_2$, however this bandgap difference between 2L and 3L is small (~100 meV), and $d_c/d_\perp$ is more influential here ($d_c/d_\perp \approx 25$ % for 2L and 40 % for 3L).

**S3. Transport and reference voltage measurements**

The $\Delta V^{\text{ref}}$ vs $\Delta V$ data in Fig. 5 of the main text was collected by applying a potential difference, $\Delta V$, directly between top and bottom gate electrodes and measuring the reference voltage of each ionic liquid (IL) with the 2DM electrically floating. Reference voltages were recorded by sourcing exactly 0 A to the reference electrodes while recording their potential (Keithley 2450). The data in Figs. 3 and 4 used top and bottom gate voltages controlled by separate sourcemeters along with a third sourcemeter to apply a drain/source potential, $V_{\text{ds}}$ (100 mV for 2L WSe$_2$ (sample #1) in Fig. 3, 50 mV for 3L WSe$_2$ (sample #2) in Fig. 4, and 100 mV for 1L WSe$_2$ (sample #4) in Fig. S3).

**S4. Spurious or leakage current in our IL-gated devices**

It is important to ensure that $I_{\text{ds}}$ in our devices is separated from the leakage current across our 2DM/IL interface. Below, we show an approach to analyze the leakage current. We model our system as a combination of several capacitors (Fig. S5): $C_b^w$ and $C_t^w$ are working electrode capacitors corresponding to the interface between bottom and top gate electrodes and ILs respectively. The counter electrode capacitors, $C_b^c$ and $C_t^c$, model the IL/2DM interface for bottom and top ILs respectively. Finally, $C_\perp$ models the direct coupling between bottom and top ILs, which couple through both the 2DM and the PMMA/SiN. We model leakage by introducing currents $I_b^w$, $I_t^w$, $I_b^c$, $I_t^c$, and $I_\perp$, each associated with their respective capacitor. These currents arise due to i) charging/discharging of the corresponding EDLs, or ii) charge transfer due to electrochemical reactions at the corresponding interface[3]. We note that when the 2DM is electrically floating, the 2DM acts as an ideal atomically thin barrier separating the two ILs. This barrier prevents electrochemical reactions between the two ILs while still enabling the study of the interaction between the two ILs when separated by an ultrathin charge-neutral membrane. Due to the expected suppression of electrochemical reactions in this state, we predict that our device geometry will provide a robust platform to study previously-inaccessible phenomena such as dielectric breakdown of the barrier material (without influence of damage from gate leakage currents) and interactions between ions at ultralow distances.

In this model, $I_b^w$ and $I_t^w$ are the currents experimentally detected by our sourcemeters applying the corresponding gate voltages. The currents $I_b^c$ and $I_t^c$ are the leakage currents flowing in the 2DM and potentially contributing to the measured drain-source current. By invoking Kirchhoff's current law for the circuit in Fig. S5, we get $I_b^w + I_t^w = I_b^c + I_t^c$. This means that we can determine the maximum amount of leakage current which could affect our measured $I_{ds}$ by simply taking the sum of leakage currents detected by the sourcemeters. The value of $I_b^c + I_t^c$ in our devices is at most a few nA, at least one order of magnitude lower than the scale of $I_{ds}$ used to obtain threshold voltages.

In our dual IL-gated FETs *without any capping layer* above the 2DM, current may flow through the suspended region of interest as well as supported regions of the flake, $I_{susp}$ and $I_{supp}$ respectively. We model this as a parallel connection of two different gate-tunable resistors, one resistor modeling the suspended dual-gated region of the 2DM (Fermi level depends on $V_b + V_t$), and the other resistor models the SiN-supported region of the 2DM (Fermi level depends on just $V_t$). The total drain-source current is $I_{ds} = I_{susp} + I_{supp}$.

We performed simple test measurements in order to confirm the PMMA capping layer's ability to neutralize the contribution of $I_{supp}$ (see Fig. S6). We fabricated a simple 2L WSe$_2$ device using the same fabrication methods outlined above, but on a different substrate (300 nm SiO$_2$ thermally grown on Si). The device has two channel regions, one without PMMA (Ch1) and one covered by crosslinked PMMA (Ch2). We apply a drop of DEME-TFSI on top of the WSe$_2$, and then we perform transport measurements of Ch1 and Ch2 simultaneously (see Fig. S6). It is clear from the data that crosslinked PMMA blocks ions from changing the carrier density in the covered region, Ch2, and given the intrinsic charge-neutrality of the WSe$_2$, we see that no current flows through the covered channel.

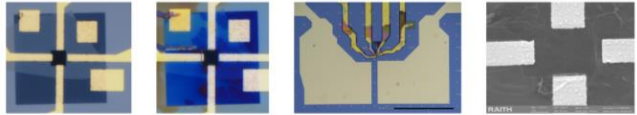

a) Sample #1: 2L WSe$_2$ w/PMMA

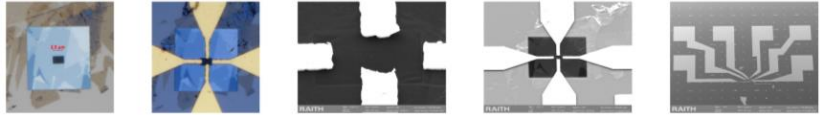

b) Sample #2: 3L WSe$_2$ w/out PMMA

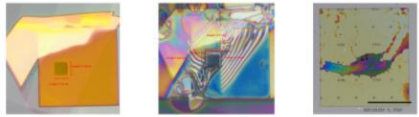

c) Sample #3: 1L WSe$_2$ w/PMMA no electrodes

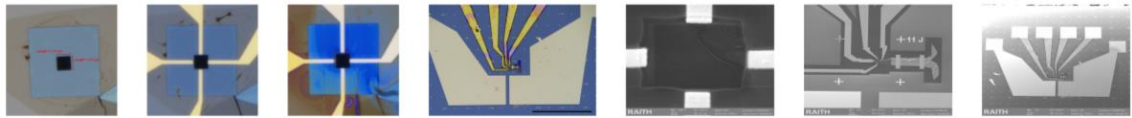

d) Sample #4: 1L WSe$_2$ w/PMMA

**Fig. S1 | Samples. a)** Sample #1 after metallization (100x), crosslinking (100x and 5x, scale bar in 5x is ~1 mm), and after measurements (SEM images). SEM images after measurements show no damage to the suspended 2DM, it remains intact and not ruptured. This sample is shown in Fig. 2b,c in the main text. **b)** Sample #2 after transfer (100x), metallization (100x), and after measurements (SEM images). SEM images after measurement show no damage to the suspended 2DM, it remains intact and not ruptured. **c)** Sample #3 after transfer (100x) and crosslinking (100x and 10x, scale bar in 10x is ~500 µm). This sample has no drain/source electrodes. **d)** Sample #4 after transfer (100x), metallization (100x), crosslinking (100x and 5x, scale bar in 5x is ~1 mm), and after measurements (SEM images).

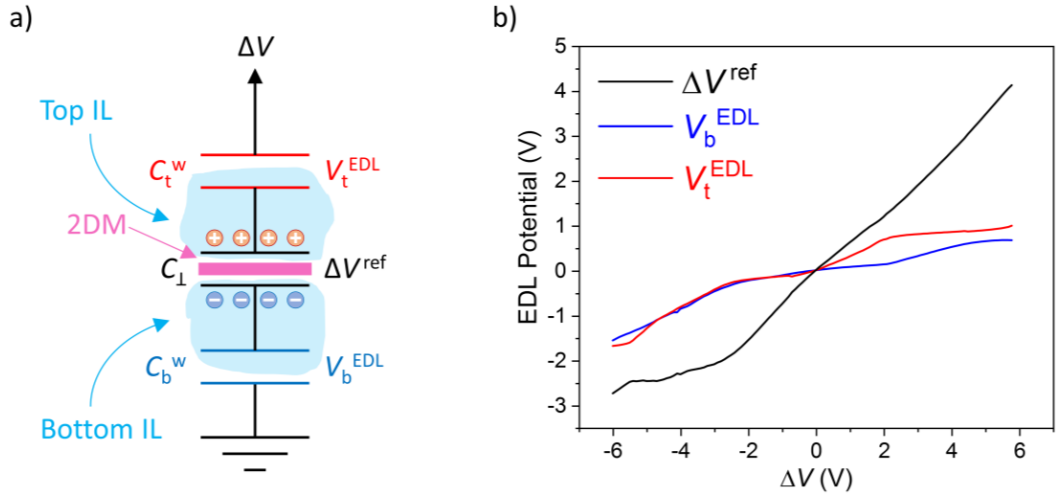

**Fig. S2 | Circuit analysis for reference voltage measurements. a)** The circuit for the reference voltage experiment in Fig. 5 of the main text (see Fig. S5 for complete circuit). **b)** Average voltage drop across the corresponding EDLs as a function of $\Delta V$ for each of the three capacitors involved (each curve vertically shifted to be centered about the origin). Note that the two gate EDLs have roughly the same value as a function of $\Delta V$, thereby confirming that $\alpha_b \approx \alpha_t$.

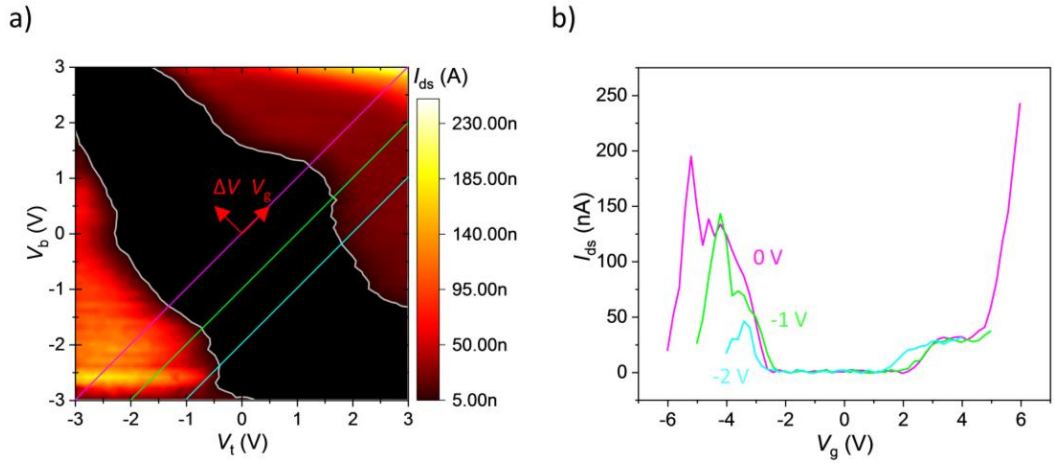

**Fig. S3 | Dual IL-gated monolayer WSe$_2$. a)** Map of $I_{ds}$ vs. ($V_b$, $V_t$) for monolayer WSe$_2$ (sample #4). **b)** Line scans corresponding to the slices of $\Delta V$ (labeled in the figure) from the map in *a*). Note how the bandgap appears to not change as a perpendicular field is applied.

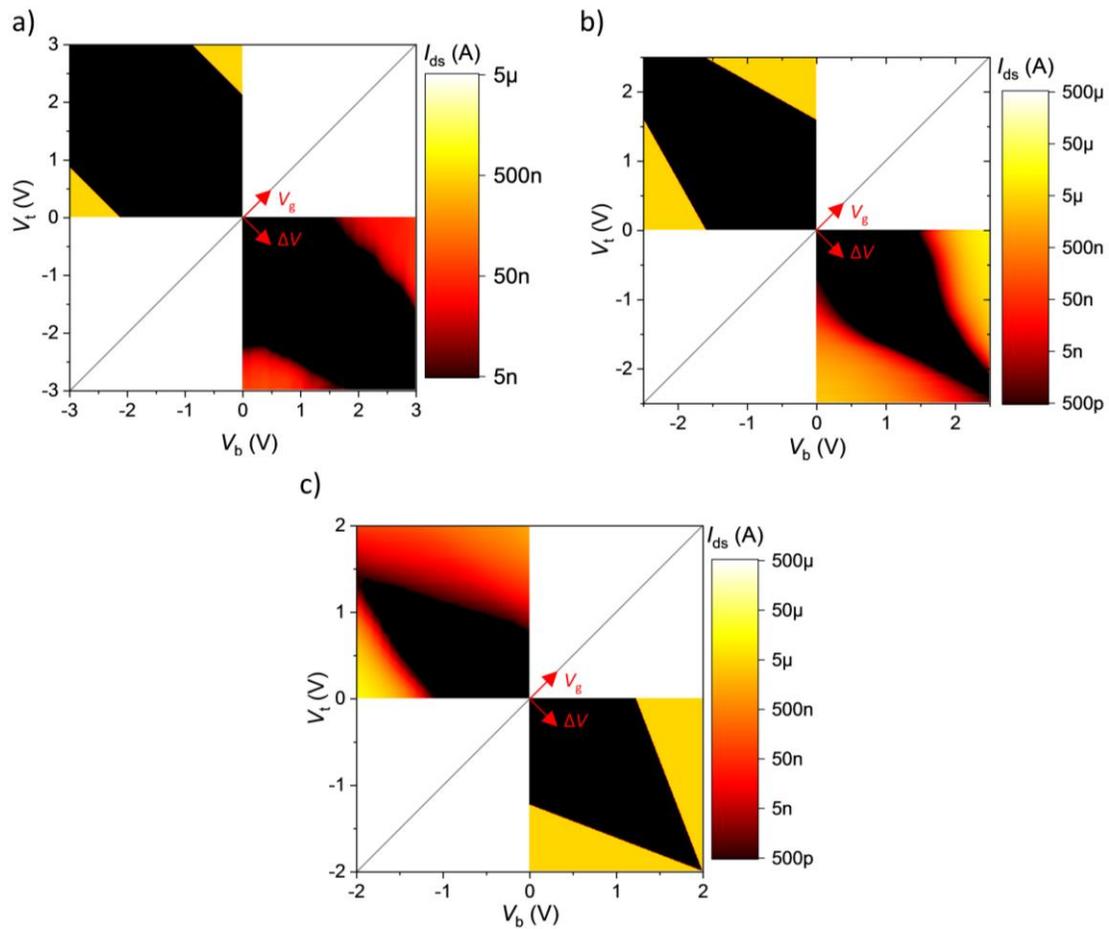

**Fig. S4 | Threshold voltage simulations. a)** Monolayer, **b)** bilayer, and **c)** trilayer WSe$_2$ transport data and threshold voltage simulations side by side in the same map. The black region of the simulation indicates the bandgap, and yellow regions of the simulation indicate electron and hole conduction. The map should be symmetric about Δ$V$ = 0 V (diagonal gray lines) due to the lack of a built-in field, enabling us to compare the transport data vs. the threshold voltages.

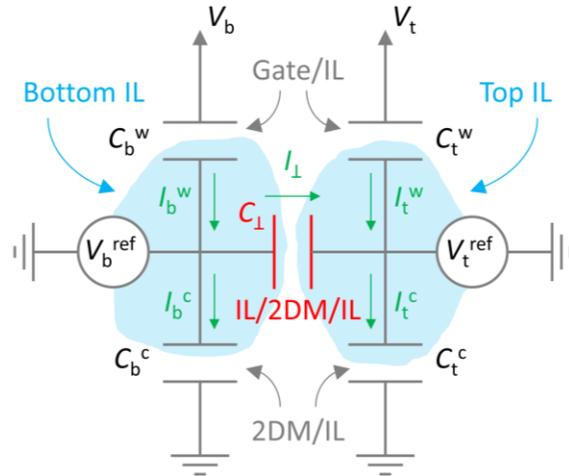

**Fig. S5 | Dual IL-gated FET gating circuit.** Dual IL-gating circuit diagram. Each individual IL has its own working (gate) electrode and its own reference voltage. When the 2DM is left floating (circuit in Fig. S2a), the bottom and top counter electrodes, $C_b^c$ and $C_t^c$ respectively, are left floating, and therefore do not affect the circuit behavior, as in the case of the reference voltage measurements using sample #3 in Fig. 5 of the main text. The leakage current potentially flowing through the device is the sum of the two counter electrode currents, which is found from the figure using Kirchhoff's current laws and summing the two measured currents from our sourcemeter: $I_b^w + I_t^w = I_b^c + I_t^c$.

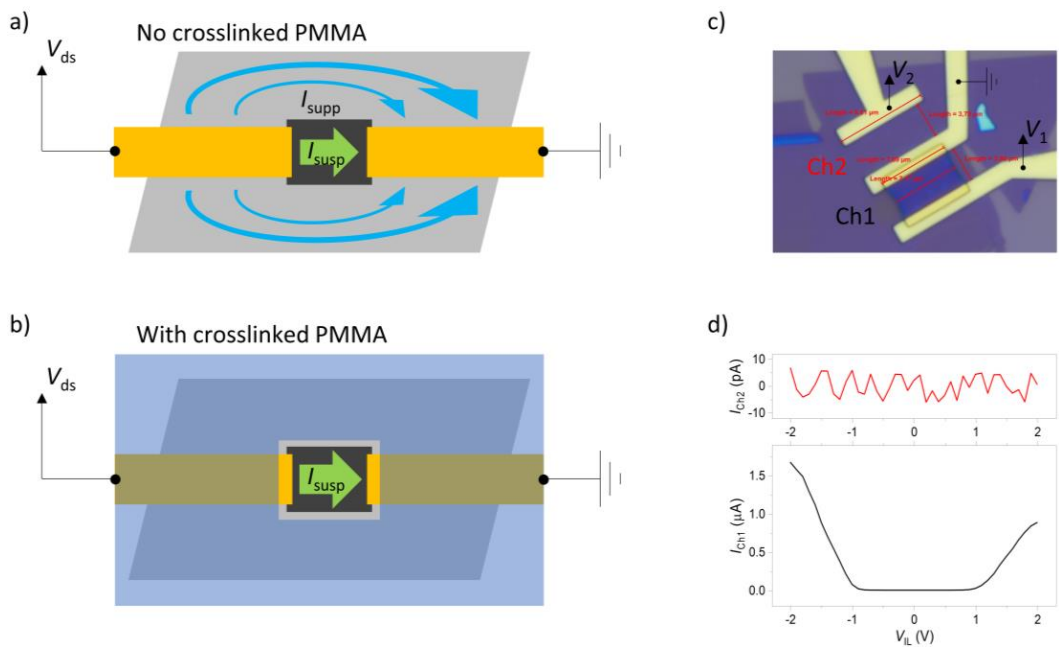

**Fig. S6 | Supported/suspended current and the need for a capping layer. a)** and **b)** show how supported current is restricted and minimized with the addition of a PMMA capping layer. **c)** 100x image (with dimensions) showing WSe$_2$ with two channel areas, uncovered (Ch1) and covered (Ch2) with crosslinked PMMA. **d)** Electrical transport through both channel regions. It is clear from the data that crosslinked PMMA blocks ions from changing the carrier density in covered regions, ensuring that no current will flow in these regions (given that the 2DM is natively charge-neutral). This also shows that our noise floor is ~10 pA, corresponding to the limits of our measurement equipment.